# A Rule of Solute Segregation at Grain Boundaries

Xin Li[1], Wang Gao[1,*], and Qing Jiang[1,*]

[1]Key Laboratory of Automobile Materials, Ministry of Education, Department of Materials Science and Engineering, Jilin University, 130022, Changchun, China.

*Emails: wgao@jlu.edu.cn; jiangq@jlu.edu.cn.

**Abstract**

The control of solute segregation at grain boundaries is essential in engineering alloy properties, however the structure-activity relationship of the key parameter—the segregation energies—still remains elusive. Here we propose the electronic and geometric descriptors of grain-boundary segregation based on the valence and electronegativity of solutes and matrices, the size of solutes and the non-local coordination number of free surfaces, with which we build a predictive analytic framework to determine the segregation energies across different solutes, matrices, grain-boundary structures and segregation sites. This framework uncovers not only the coupling rule of solutes and matrices, but also the origin of solute-segregation determinants. The contribution of solutes stems from the d- and sp-state hybridization in alloying, whereas the geometric contribution of matrix grain-boundary interfaces is determined by matrix free surfaces. Our scheme builds a novel picture for grain-boundary segregation and provides a useful tool for the design of advanced alloys.

**Teaser**

The structure-activity relation of segregation energies is proposed across different solutes, matrices, grain boundaries and segregation sites.

**Introduction**

Solute segregation at grain boundaries (GBs) plays an essential role in governing the mechanical, functional, and kinetic properties of metallic materials because it controls the chemical composition and local structure of GBs(*1-4*). This critical segregation behavior can be quantitatively characterized by the segregation energies, which measure the preference of solute atoms to segregate to GBs or remain inside the bulk. Generally, the segregation energies depend on the electronic and geometric properties of solutes and matrices. However, it still presents a fundamental challenge to rapidly evaluate the GB segregation energies and to understand the underlying mechanism of solute segregation.

While many studies of segregation energies focus on the numerical calculations of individual systems using density functional theory (DFT) and molecular dynamics (MD) methods(*5-12*), some frameworks have been proposed for predicting the segregation energies. Early empirical thermodynamic models attempted to determine the segregation energies by using macroscopic properties, such as the bulk modulus and shear modulus in the McLean model(*13*) and the solution enthalpy and solute strain energy in the Miedema model(*14, 15*). However, these models rarely explain the electronic origin of GB segregation and are rather limited in predicting the segregation energies(*16, 17*). Some machine learning methods can be used to predict the segregation energies accurately(*18, 19*), but generally cannot provide a clear physical picture of segregation behaviors. Other effective descriptors were also proposed to determine the segregation energies, such as the parameters based on the d-band properties(*20-22*), excess volumes(*23*) and coordination numbers(*18*). Essentially these descriptors characterize the structures of matrices, however, none of them can reflect the electronic effect of solutes. In addition, some descriptors like the d-band properties and excess volumes, derived from DFT calculations, are time-consuming to obtain and even with some numerical uncertainties. Moreover, they are hardly applicable to complicated GBs such as those in polycrystals.

Here, we find that the valence and electronegativity of solutes and matrices and the size of solutes combined together with the generalized coordination number of segregation sites control the segregation energies. With these easily accessible parameters, we build a predictive analytic framework for evaluating the segregation energies across the different solutes, matrices, GBs and segregation sites. This framework reveals that the solute effect of GB segregation originates from the solute d- and sp-state hybridization with neighboring atoms in alloying and the matrix interface geometric effect from the matrix free surfaces. These results provide a deep understanding of the GB-segregation mechanism and a physical guidance for designing the alloy structures with targeted properties.

**Results**
**The effect of different solutes in a given matrix.** First, we focus on the role of different transition-metal (TM) solutes in GB segregation for a given matrix. Generally, the shift of a solute atom from the matrix to GB involves the bond breaking and forming and is limited by the volume of segregation sites. We thus attempt to describe the segregation behaviors from the alloying effect between solute and matrix (the interatomic bonding effect of alloying atoms) and from the size of solutes. The tight-binding (TB) and Friedel's models(*24, 25*) have revealed in TMs that the d-band width $W_d$ (defined as the second moment of the projected d-band density of states(*26*)) is dominant in the formation of the cohesive force and the s-band contribution dictates the mole volume and compressibility(*27*). It is known that $W_d$ depends strongly on the valence-electron number ($S_v$) and the interatomic bond strength does on modified Pauling electronegativity ($\chi$)(*28*). We thus adopt $\psi = \frac{S_v^2}{\chi}$, which can describe the surface bond breaking reasonably(*29, 30*), to study the segregation energies ($E_{seg}$). $\psi$, which is a rough piecewise linear function of $W_d$ and cohesive energy (see Fig. S1), provides a rough description of the segregation energies in the W matrix (the structures are shown in Fig. S2). According to the TB models, the alloying and size effects of solutes contain the d-d state and d-sp state hybridization of the neighboring atoms as well as the charge transfer ($\Delta Q$) between the solute and matrix atoms. It is essential to quantify these effects on the d-band width projected on the solute atoms $W_d^{sol}= f (V_{dd}, V_{d\text{-}sp}, \Delta Q)$ in order to determine $E_{seg}$ (where $V$ is the interatomic matrix element). The d-d orbital bonding interaction can be reflected by the interatomic matrix element ($V_{dd}$) as follow:

$$V_{dd}^{(i,j)} = \eta_{dd} \frac{\hbar^2 [r_d^{(i)} r_d^{(j)}]^{\frac{3}{2}}}{m d_{i,j}^5} \tag{1}$$

where $d_{i,j}$ represent the internuclear distance between $i$ and $j$ atoms. $m$ is the mass of an electron. $r_d$ is spatial extent of the d-orbitals. $\eta_{dd}$ is a dimensionless proportionality constant that is formally related to the type of d-d orbital bonding interaction. Thus, the d-d state hybridization is proportional to $W_d^{sol}$ that depends on the bond distance $d$ in the power law of $1/d^5$ in Ref. [(*31*)], leading to the change of $W_d^{sol}$ as a function of the local geometry $\Delta W_d^{sol} = f(V_{dd}) \propto 1/d^5$. The second kind of interactions stems from the d-sp orbital hybridization and the charge transfer between the solute and matrix atoms. Xin et al have reported that the electronegativity difference after alloying can accounts for the interaction between local d-states and the sp-states of neighboring atoms as well as the charge transfer(*32*). Thus, the corresponding change of the d-band width or d-band center after alloying can be obtained by introducing the electronegativity difference with a constant parameter. The atoms' electronegativity is proportional to the bond distance $d$ in Ref. [(*32*)], the rough linear relation between $\psi^{sol} = \frac{S_v^2}{\chi}$ and $W_d^{sol}$ indicates $\Delta W_d^{sol} \propto f(V_{d\text{-}sp}, \Delta Q) \propto 1/\Delta\chi \propto 1/d$. Therefore, we approximately use the $\Delta W_d^{sol} \propto f (V_{dd}, V_{d\text{-}sp}, \Delta Q)= f (V_{dd}) f (V_{d\text{-}sp}, \Delta Q) \propto V \propto 1/d^6$ to reflect the alloying and size effects of solutes (the attractive interaction) during GB segregation.

According to the TB models, the interaction between two atoms contains the bonding and anti-bonding states, corresponding to the attraction ($V$) and overlap (Pauli repulsion, $S$) for the electrons. $S$ is proportional to $V$ and the bond energy can be expressed as $\Delta E=2S|V|-2|V|=2\alpha V^2-2|V|$ as demonstrated in Ref. [(33)], in which the repulsive interaction ($V^2$) is approximately the square of the attractive interaction ($V$) and thus decays as $1/d^{12}$. The interatomic distance $d$ depends on the radius of solute atoms ($R$) and the alloying and size effects of solute segregation are proportional to the term of $[\left(\frac{\sigma}{R}\right)^{12} - \left(\frac{\sigma}{R}\right)^{6}]$. Therefore, we introduce a Lennard-Jones (L-J) potential like formula to determine $W_d^{sol}$ and the solute segregation energies as follows,

$$D_{seg}=[\left(\frac{\sigma}{R}\right)^{12} - \left(\frac{\sigma}{R}\right)^{6}]\times\psi^{sol} = [\left(\frac{\sigma}{R}\right)^{12} - \left(\frac{\sigma}{R}\right)^{6}]\times\frac{S_v^2}{\chi} \qquad (2)$$

Here, $R$ is the radius of an atom X which is defined as half the length of a X-X single bond in a crystal or molecule. $\sigma$ with a certain range of 1.18~1.21 has a minor effect in determining the segregation energies and thus is taken as the constant 1.20 for all considered matrices. All the parameters in $D_{seg}$ can be obtained by table looking up (Table S1).

We first study the solute segregation at the ideal W GBs, by adopting 8 symmetric tilt GB structures ($\Sigma$3(111), $\Sigma$3(112), $\Sigma$5(210), $\Sigma$5(310), $\Sigma$7(213), $\Sigma$9(114), $\Sigma$11(323) and $\Sigma$13(510)) with five different substitutional sites along each GB structure. The symmetric GBs are formed by the two identical free surfaces rotating at a certain angle around the rotation axis, and the ideal GBs experience no sliding or migration when the solute atoms segregate. The ideal GBs and substitutional sites are illustrated in Fig. 1 and Fig. S3. The results show that $E_{seg}$ exhibits a broken-line relationship with $D_{seg}$ for the considered TM solutes with reasonable accuracy regardless of the GB structures and segregation sites (Fig.1 and Table S2). The turning point of the broken-line scaling is around $D_{seg}$ = -5 (Cr and V) and $D_{seg}$ = -10 (Hg and Au), dividing the solutes into the three groups.

These results essentially originate from the different d- and s-state properties of TM atoms in alloying. The TB models show that $W_d^{sol}$ dominates the trend of the cohesive force of IIIB-VIII group TMs and the s-bands dictate that of the IB and IIB group TMs(27). The segregation behavior of the solutes in groups 1 and 2 follows the trend of the cohesive energy of IIIB-VIII group TMs, reflecting the dominant role of the d-states in alloying IIIB-VIII group solutes with W. On the other hand, the behavior of the solutes in group 3 is opposite to that in group 2, corresponding to the dominant role of the s-states in alloying IB-IIB group solutes with W.

The size effect is secondary compared to the electronic bonding of solutes, in stark contrast with the viewpoints that proposed the solute-size effect to play a major role in GB segregation(5, 6). A L-J potential like formula(34) of $D_{seg}$ implies that the size of solutes exhibits mainly a two-body effect in GB segregation with negligible many-body effects. As $\psi^{sol}$ reflects $W_d^{sol}$ and s-band filling and the size of solutes does significantly the s-band depth (which is the energy of the s-band bottom)(27), the descriptor $D_{seg}$ uncovers the coupling nature between the d- and s-states of solutes in determining the segregation energies.

For the three group solutes in the W matrix, the corresponding slopes and intercepts of the linear scaling are labeled with $k_i$ and $b_i$ ($i$=1-3). Notably, $k_1$, $k_2$ and $k_3$ are linearly dependent on each other for all the GBs and segregation sites (see Fig. 2a and b), and so are $b_1$, $b_2$ and $b_3$ (see Fig. 2c and d). Interestingly, for each solute group, the slope is also a linear function of the intercept, such as $k_2$ and $b_2$ (Fig. 2e). These results indicate a sole physical origin for all slopes and intercepts (that will be further discussed below).

We now consider the segregation energies in W with GB relaxation during solute segregation. For $\Sigma$5(210), $\Sigma$7(213), $\Sigma$9(114), $\Sigma$11(323) and $\Sigma$13(510) GBs, they experience GB sliding or migration and lose their mirror symmetry after inserting the solute atoms(5, 35, 36) (see Fig. S5). Nevertheless, $D_{seg}$ is still able to determine the trends of the segregation energies for various GBs and segregation sites (see Fig. S6 and Table S3), further indicating that $D_{seg}$ reflects the core elements of the alloying and size effects of TM solutes for GB segregation.

In the case of other metal matrices, such as Nb, V, Ir, Rh, Os, Tc (the structures are illustrated in Fig. S4), Al(*37*), Ni(*38*), Zr(*39*) and Mo(*17*), we find that $D_{seg}$ always exhibits a broken-linear relationship with $E_{seg}$ (Fig. 3, Figs. S7 and S8 and Tables S4-S6). These broken-line relationships lead to the $d^5$ state at the turning point of groups 1 and 2 and the $d^{10}$ state at that of groups 2 and 3. For the TM matrices, the solute itself exhibits this feature. For the main-group matrices such as Al, an Al atom has three valence electrons and forms the $d^5$ state with Ti ($3d^2$) and the $d^{10}$ state with Co ($3d^7$) at the turning points. Importantly, this feature is crucial for the prediction of segregation energies.

Furthermore, $D_{seg}$ is also applicable to the solute segregation of twist GBs and even polycrystal GBs (that are the irregular grain boundaries between multiple grains with different orientations in polycrystals illustrated in Fig. S9)(*19*). $D_{seg}$ exhibits a broken-line relationship with $E_{seg}$ in Mo Σ5(100) twist GB with the mean absolute error (MAE) of 0.13 eV at site 1 and 2 (Fig. 3d and Fig. S8d). More importantly, it can determine the trend of segregation energies for a given segregation site in polycrystals of 15 matrices such as Ti, Zr, Ta, Co, Fe, Ni, Mo, W, Pd, Pt, Au, Ag, Cu, Mg and Al, with the accuracy of 0.05 eV (see Figs. S10-S12 and Table S7), comparable to that of the smooth overlap of atomic positions (SOAP) based machine learning work(*19*). The deviation of some data relative to the fitting lines is likely due to the fact that all data at polycrystal GBs are obtained with MD calculations (which generally bring out larger errors compared with those obtained by DFT calculations). Interestingly, some main-group elemental solutes also follow the $D_{seg}$-determined relations of TM solutes, indicating that the valence origin of $D_{seg}$ is useful to understand the behavior of all kind solutes (see Fig. S8b).

Overall, all the above findings demonstrate that the descriptor $D_{seg}$ quantifies effectively the alloying and size effects of solutes at the GB segregation across different kind GBs and matrices.

**The geometric effects of matrix GBs.** We now attempt to understand the geometric effects of GB structures in segregation behavior, which corresponds to the various GB interfaces and the environment of segregation sites. These effects are influenced by many factors such as GB interface sliding and migration as well as the local deformation. We start with the simple ideal symmetric tilt GBs to discuss the geometric effects of segregation energies. Since the symmetric tilt GBs are formed by two identical grains (or free surfaces), we use the generalized coordination numbers ($\overline{CN}$) of the unrelaxed free surfaces (before GB formation) without any solute and vacancy to characterize GB geometric properties. The expression of $\overline{CN}$ is as follows:

$$\overline{CN}^{(m)} = \frac{\sum_{i=1}^{n} \overline{CN}^{(m-1)}(i)}{CN_{max}} \tag{3}$$

Here, $i$ represents the $i$th neighbor of a given surface atom. $n$ is the number of neighboring atoms. The unrelaxed free surfaces do not experience any change of bond length and bond angle, and thus the identification of neighbors for a given surface site does not need a cutoff radii, which avoids the numerical uncertainty. $CN_{max}$ is the max coordination number in the bulk. $m$ denotes the order of approximation. The ordinary coordination number $CN$ corresponds to the zeroth order approximation $\overline{CN}^{(0)}$, reflecting the geometric environment of a surface atom with its nearest neighbors. We use the first-order approximation $\overline{CN}^{(1)}$, that is, the effect of second-neighboring atoms is included. Notably, this approximation was first proposed to determine the adsorption of small molecules on TM surfaces and nanoparticles(*40, 41*). If one adopts the second- or higher-order approximation of $\overline{CN}^{(m)}$ ($m \geq 2$), the effects of farther neighboring atoms are involved ($\overline{CN}^{(1)}$ and $\overline{CN}^{(2)}$ are shown in Table S8).

Fig. 4(a-c) shows that for a given solute atom segregated in the ideal W matrix, $E_{seg}$ is a broken-line function of the $\overline{CN}^{(1)}$ of the segregation sites on the unrelaxed free surfaces for each GB.

Moreover, most of the segregation sites comply with the same $\overline{CN}^{(1)}$-determined relation. In contrast, the ordinary *CN* (namely $\overline{CN}^{(0)}$) fails to determine the segregation energies (Fig. S13). $\overline{CN}^{(2)}$ is almost the same as $\overline{CN}^{(1)}$ in determining $E_{seg}$ (see Fig. S14), indicating that the geometric effect of a segregation site is mainly from its nearest and second-nearest neighbors with a negligible contribution from farther neighbors. The turning point of the broken-line scaling is around $\overline{CN}^{(1)}=$ 4, which is roughly the boundary between site 1 and site 2. The reverse folding of site 1 likely stems from the fact that the structure of site 1 at GBs is close to that of the bulk and is thus different from that of surfaces. Besides the ideal W matrix, the $\overline{CN}^{(1)}$ of the unrelaxed free surfaces is also able to describe the effect of different segregation sites for the GBs with multiple metastable structures, such as reflection and isosceles of {112} twin boundaries(*44*) in the W matrix (see Fig. S15). Although these two GB structures exhibit two different sets of segregation energies for a given solute (such as Ir, Pd or Pt), each set can be determined accurately by the same $\overline{CN}^{(1)}$ (see Fig. S15). The different interface structures are reflected by the slopes and intercepts of the $\overline{CN}^{(1)}$-determined scaling. These findings also hold for the other five GBs (with both symmetric and asymmetric structures) in the W matrix, such as Σ5(210), Σ7(213), Σ9(114), Σ11(323) and Σ13(510) GBs (Fig. S16). Notably, the segregation site 1 of the relaxed GBs generates two linear relations, mainly caused by the different structural deformations of various GB structures: one is for the ones with GB sliding and the other is for the ones without GB sliding. These two lines are also predictive, if both are included, the MAEs of the predicted segregation energies are decreased from ~0.25 eV to ~0.18 eV. The prediction error of $\overline{CN}^{(1)}$ is around 0.2 eV that is within that of the DFT semi-local functional calculations (*42, 43*). Clearly, the accuracy of our scheme is usually better than 0.2 eV (which is the upper limit of the average error in our study) and the prediction error is smaller than 10% of the range of the segregation energies. Considering that the range of the studied segregation energies is ~2.5 eV for the ideal GBs and ~6 eV for the relaxed GBs (Fig. S16), the 0.2 eV error is sufficient for the determination and prediction of the segregation energies. In particular, our scheme is also effective for the segregation with low segregation energy: the 0.03 eV error for the Mo solute segregating in W matrix GBs with the range of segregation energies of ~0.30 eV (Fig. S13e) and the 0.01 eV error for the Ir, Pd or Pt solutes segregating in the twin boundaries of W matrix with the segregation-energy range of ~1.5 eV (Fig. S15). These results imply that the geometric contribution of matrix GBs to the segregation energies is non-local compared with the effect of solutes and mainly determined by the unrelaxed matrix free surfaces regardless of the GB deformation. The non-local $\overline{CN}^{(1)}$ can be used to quantify the geometric effect of the segregation energies for both the symmetric and asymmetric GBs formed by the low-index and high-index surfaces.

Our findings are also comparable to the literature results. Hu *et al.* considered the geometric effect of W and Ta matrix GBs with the linear combination of the bimodality property of the d-bands and the strength of sp–d hybridization in the GB structures (that are obtained by DFT calculations)(*20*). $\overline{CN}^{(1)}$ is found to linearly depend on the linear combination of their descriptors (see Fig. S17), supporting the effectiveness of $\overline{CN}^{(1)}$ in evaluating the GB segregation. Huber *et al.* use the excess volume and ordinary *CN* to determine the geometric effect of segregation sites and the excess volumes are calculated from the relaxed GB structures with DFT calculations(*18, 23*). In comparison, $\overline{CN}^{(1)}$ originates from the unrelaxed free surfaces and is independent of DFT calculations, avoiding the numerical uncertainty.

**The effect of a given solute in different matrices.** To exclude the effect of lattice structures, we study the segregation energies of a given solute in 8 body-centered cubic (bcc) (Cr, Mn, Fe, V, W, Nb, Mo and Ta) and 7 face-centered cubic (fcc) (Ag, Au, Cu, Rh, Ir, Pd and Pt) GBs. The corresponding segregation energies in different matrices are the broken-line function of the $\psi^m$ of matrices at one segregation site for one lattice structure (see Fig. S18). This reflects the fact that the solute atoms only have a negligible effect on the d-band width of matrices due to their low concentration and local effects, which is different from the effect of matrices on the d-band width projected on solute atoms $W_d^{sol}$. The d-band width of matrices thus dominates the types (electronic effects) of matrices on GB segregation. Therefore, the segregation energies of different solutes and matrices at each segregation site can be determined by the linear combination of the $D_{seg}$ of solutes and the $\psi^m$ of matrices (Fig. S19 and Table S9). The slopes and intercepts of these broken-line relationships (as well as those of the scaling between $E_{seg}$ versus $D_{seg}$) can be characterized as the linear function of $\overline{CN}^{(1)}$ of segregation sites (Fig. 4d and Fig. S20), which reveals a coupling mechanism between the solutes and matrices.

**A general scheme to determine the segregation energies.** Therefore, we can propose an entire expression to determine the segregation energies, by combining the $D_{seg}$ of solutes, the $\psi^m$ of matrices, and the $\overline{CN}^{(1)}$ of segregation sites together. For the group 2 solutes segregating to the sites 2-5 of bcc matrices (W, Mo, Cr, Mn and Fe), the expression is as:

$$E_{seg} = (-0.08\overline{CN}^{(1)} + 0.67)(D_{seg} + 0.33\psi^m - 1.33) \tag{4}$$

For bcc V, Nb and Ta matrices, $E_{seg} = (0.015\,\overline{CN}^{(1)} - 0.155)(D_{seg} + 0.33\psi^m - 3)$. According to the linear correlations of the slopes and intercepts for the three solute groups, one can determine $E_{seg} = -k_2(2D_{seg} + 0.33\psi^m - 1.33)$ for the group 1 solutes and $E_{seg} = -k_2(0.33D_{seg} + 0.67\,\psi^m - 2.67)$ for the group 3 solutes in the W matrix (here, $k_2 = -0.08\,\overline{CN}^{(1)} + 0.67$). The expression of segregation energies for fcc matrices are shown in Supplementary Text.

Our scheme uncovers a unique and simple physical picture for the coupling between solutes and matrices. Moreover, the involved parameters $S_v$, $\chi$, $R$ and $\overline{CN}^{(1)}$ are easily accessible by table looking up and thus convenient for the practical application. Overall, our descriptors and models hold for the segregation of 49 solutes at 24 GBs of 21 matrices (totally 6431 data): the 29 TM solutes and 20 main-group solutes; the bcc (W, Mo, Ta, Nb, V, Cr, Mn and Fe), fcc (Ag, Au, Cu, Pd, Pt, Ni and Al) and hexagonal close-packed (hcp) (Os, Tc, Ti, Zr, Co and Mg) matrices. These include the major binary alloy systems of interest in the field. The studied 24 GBs contain the ideal symmetric tilt and twist GBs, the asymmetric GBs with sliding and migration and even complicated polycrystal GBs. The MAE of prediction is 0.10 eV in 8 W GBs and Ir and Os GBs, 0.05 eV for 2 Nb GBs, 0.06 eV for V GBs, 0.09 eV for Rh and Tc GBs, 0.11 eV for 12 Al GBs, 0.16 eV for Ni GBs, 0.14 eV for Mo GBs and 0.14 eV for Zr GBs and 0.05 eV for polycrystals of 15 matrices. These results demonstrate that our scheme is universal and predictive in determining the segregation energies across different TM solutes, matrices, GB structures and segregation sites.

Our scheme can also deduce or rationalize the experimental segregation tendency. The experimental results in W show that there is almost no segregation for Ti or Ta solutes at GBs, whereas Ag solute is prone to segregate to all sites around GBs(*45*). These observations likely stem from the fact that the Ti/Ta and Ag solutes sit close to the turning points of our scheme, the top and bottom respectively. More importantly, combining our scheme with the White-Coghlan model(*46*) (which exhibits a better description of the GB enrichment with temperatures by a distribution of segregation energies, compared to the Langmuir-McLean isotherm relationship by a single per-solute value(*18*)), one can also predict the GB solute concentration ($c_{GB}$) curves for non-interacting solutes in the W matrix, which is determined as:

$$c_{GB} = \frac{1}{N}\sum_{i=1}^{N}\frac{1}{1+\frac{1-c_{bulk}}{c_{bulk}}\exp(-\frac{E_{seg}^{X,i}}{k_B T})} \tag{5}$$

where $N$ represents the total number of segregation sites (fixed as 5 here), $E_{\text{seg}}^{X,i}$ is the segregation energy of solute $X$ segregated to the position $i$ at GBs and $T$ is temperature. $c_{\text{bulk}}$ is the solute concentration in the matrix, which is fixed as 2 at.% here(*18, 20*). Fig. 5 and Fig. S21 show that the predicted GB solute concentrated curves based on our framework are close to the DFT-calculated ones across a wide temperature range regardless of solutes and GB structures, further verifying the validity of our framework. Compared with the late TM solutes, the early TM solutes have the solute concentrate depending more strongly on temperature in the W matrix, suggesting temperature as an effective way to engineer the early TM solutes at GBs.

**Discussion**

Overall, our framework provides an effective tool to analyze the solute segregation at GBs with the easily accessible intrinsic properties. The geometric effect of matrices in GB segregation can be determined by only using our proposed generalized coordination number but not the previously suggested ordinary coordination numbers (Fig. 4 and Fig. S13). Our electronic descriptor $\psi^m$ can quantify the effect of matrix types in the segregation energies effectively. The previous studies have shown that the size of solutes cannot determine the segregation energy alone(*5, 6*), nor can the electronegativity and valence-electron number of solutes. Accurate prediction of the segregation energy can only be achieved through the combination of these three parameters into a descriptor $D_{\text{seg}}$ (that accounts for the alloying and size effect of solutes), as we found. Our scheme identifies the geometric and electronic descriptors for matrices and a single descriptor for solutes and uncovers the coupling rule between the different properties of matrices and solutes with simple and general equations. The proposed descriptor $D_{\text{seg}}$ essentially implies the local effects of solute atoms and the similar matrix deformation with different solute atoms, whereas $\overline{CN}^{(1)}$ reflects the non-local effects of segregation sites at GBs and is well suited to single-crystal matrices such as the symmetric and asymmetric tilt GB structures. For polycrystal matrices, one may introduce more complicated geometric features such as bond length and angle to characterize the geometric structures of GBs, which will thus be a natural extension of this study.

In summary, we have identified the main factors that control the segregation energies of solutes at GBs: the valence and electronegativity of solutes and matrices, the size of solutes and the non-local coordination number of free matrix surfaces. This enables us to build a predictive analytic framework to quantitatively determine the segregation energies, which holds across the different solutes, matrices, GBs and segregation sites. The determinant of solutes, initially inspired by the TB models, reflects the hybridization of d- and sp-states of solutes and neighboring atoms in alloying, whereas the geometric determinant of GB matrices, indicating a non-local geometric effect, does the free-surface-determined nature of GB interfaces. Our framework uncovers the electronic origin for the behavior of alloying elements in GB segregation, the connection between free surfaces and their resulting GB interfaces, and the coupling mechanism of solutes and matrices, all of which propose a simple and clear physical picture for the solute segregation at GBs. All these findings are thus expected to serve as quantitative guidelines for the future alloy design, particularly considering that all involved parameters are predictable.

**Method**

Our calculations are performed using Vienna ab-initio simulation package (VASP) code(*47*). The interactions between ions and electrons are described by the projector augmented wave potential (PAW) method(*48*). The Perdew-Burke-Ernzerhof (PBE)(*49*) exchange-correlation functional is adopted for all the studied systems unless otherwise stated. The cutoff energy of 500 eV is used in the calculations and the k-point mesh is given in Table S10. The convergence tests of the cutoff energy, k-point mesh and functionals are summarized in Fig. S22. The cell volume and atomic position are relaxed for each GB supercell without solutes and then kept fixed when the solutes are inserted. The negligible difference between the segregation energies of the relaxed and

Page 7 of 13

fixed cell can be found in Table S11. For the optimization, all the atoms are relaxed until the force on each of them is less than 0.02 eV/atom in our calculations.

The symmetric tilt grain boundaries are constructed by using the coincidence site lattice (CSL) model(*50*). We construct 8 low-Σ symmetric tilt W GBs with [001], [110] and [111] tilt axes. The slabs are separated with 10 Å of vacuum to exclude the interactions between periodic images. The grains that are used to construct the W GBs contain the six layers for Σ3(111), Σ3(112), Σ5(210) and Σ5(310) GBs, and nine layers for Σ7(213), Σ9(114), Σ11(323) and Σ13(510) GBs. The fcc matrix Σ5(310) GBs contain the six layers and the hcp matrix {11$\bar{2}$1} twin boundaries contain nine layers. These GBs are modeled with 2×2×1 supercells. The tilt angles, layers, atom numbers and lattice parameters are also listed in Table S10. We also study the 8 tilt W GBs with structural optimization during solute segregation, the data of which are from Ref. [5] (with the exchange-correlation functional of Perdew-Wang (PW91)(*51*)). Moreover, we study the segregation energies in the GB structures of Zr(*39*), Ni(*38*), Mo(*17*) and Al(*37*). The GBs of the polycrystal Ag, Au, Cu, Fe, Co, Ni, Mo, W, Pd, Pt, Ti, Zr, Ta, Mg and Al matrices are also considered for segregation(*19*).

The segregation energy $E_{seg}$ is defined as:

$$E_{seg} = E_{GB} - E_{Bulk} \qquad (6)$$

where $E_{GB}$ is the total energy of the structure with the solute atoms at GBs and $E_{Bulk}$ is the total energy of the structure with the solute atom inside the bulk.

**Acknowledgments**

**Funding:** The authors are thankful for the support from the National Natural Science Foundation of China (Nos. 22173034, 11974128, 52130101), the Opening Project of State Key Laboratory of High Performance Ceramics and Superfine Microstructure (SKL201910SIC), the Program of Innovative Research Team (in Science and Technology) in University of Jilin Province, the Program for JLU (Jilin University) Science and Technology Innovative Research Team (No. 2017TD-09), the Fundamental Research Funds for the Central Universities, and the computing resources of the High Performance Computing Center of Jilin University, China.

**Author contributions:** W.G. and Q.J. conceived the original idea and designed the strategy. X.L. performed the DFT calculations. W.G. derived the models and analyzed the results with the contribution from X.L. X.L. and W.G. wrote the manuscript. X.L. prepared the Supplementary Information and drew all figures. All authors have discussed and approved the results and conclusions of this article.

**Competing interests:** The authors declare no competing interests.

**Data availability**
All the data related to this work are available from the authors upon reasonable request.

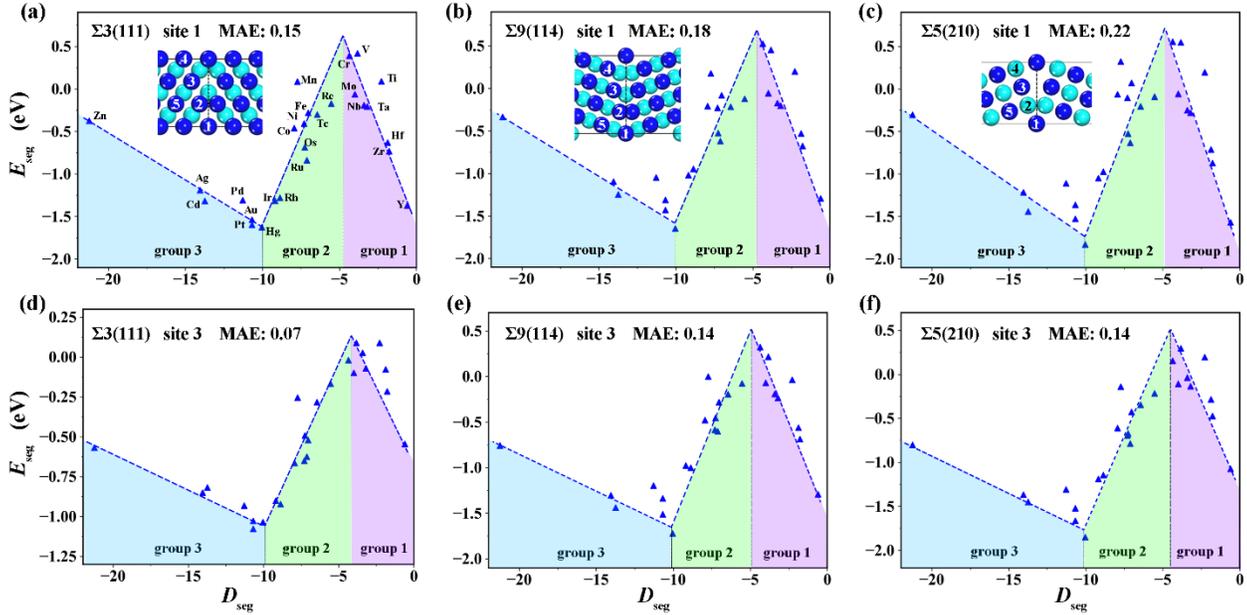

**Fig. 1 The segregation energies of solutes as a function of the descriptors $D_{seg}$ at different W grain boundaries (GBs) and segregation sites.** (**a**) The segregation site1 of Σ3(111) GB, (**b**) The segregation site1 of Σ9(114) GB, (**c**) The segregation site 1 of Σ5(210) GB, (**d**) The segregation site3 of Σ3(111) GB, (**e**) The segregation site3 of Σ9(114) GB, and (**f**) The segregation site3 of Σ5(210) GB. The structures of the three GBs and corresponding substitutional sites are illustrated in the subgraphs of (a)-(c).

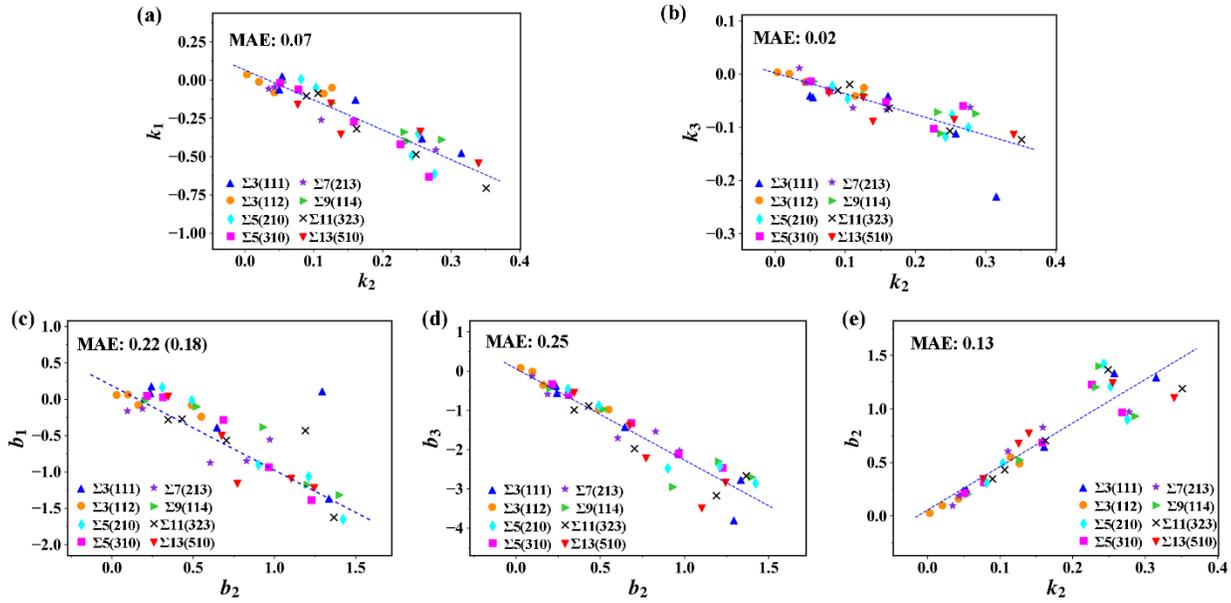

**Fig. 2 The linear scaling relationship of slopes and intercepts of different transition-metal solute groups.** (**a**) The relationship of slopes $k$ between group 1 and group 2. (**b**) The relationship of slopes $k$ between group 3 and group 2. (**c**) The relationship of intercepts $b$ between group 1 and group 2. The mean absolute error (MAE) in parentheses is the accuracy without the two outliers. (**d**) The relationship of intercepts $b$ between group 3 and group 2. (**e**) The relationship between the slopes $k$ of group 2 and the intercepts $b$ of group 2.

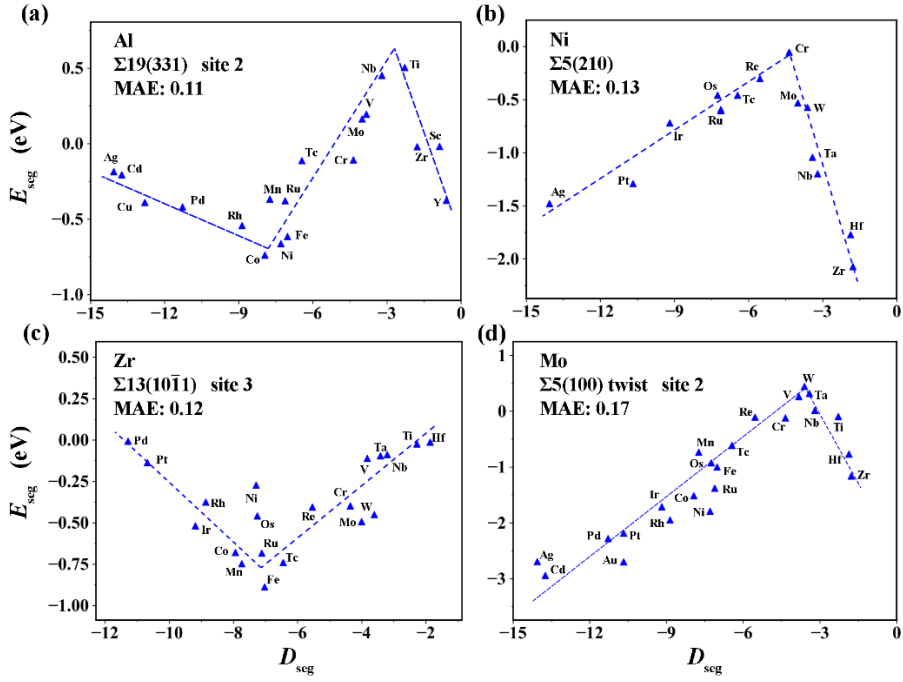

**Fig. 3 The segregation energies of solutes as a function of the descriptors $D_{\text{seg}}$ for various matrix grain boundaries (GBs).** (**a**) the Al Σ19(331) GB[36], (**b**) the Ni Σ5(210) GB[37], (**c**) the Zr Σ13(10$\bar{1}$1) GB[38] and (**d**) the Mo Σ5(100) twist GB[17].

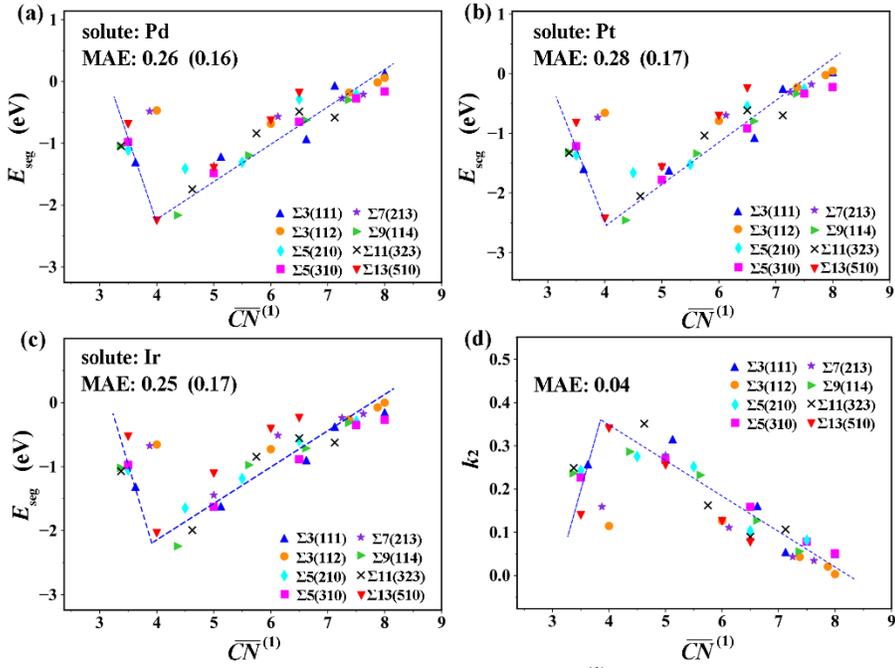

**Fig. 4 The segregation energies and the slope $k_2$ as a function of $\overline{CN}^{(1)}$.** (**a**) Pd solute, (**b**) Pt solute and (**c**) Ir solute. The mean absolute error (MAE) in parentheses is the accuracy without the two outliers. (**d**) The slope $k_2$ of the $D_{\text{seg}}$ determined scaling relation against $\overline{CN}^{(1)}$ for all the involved segregation sites.

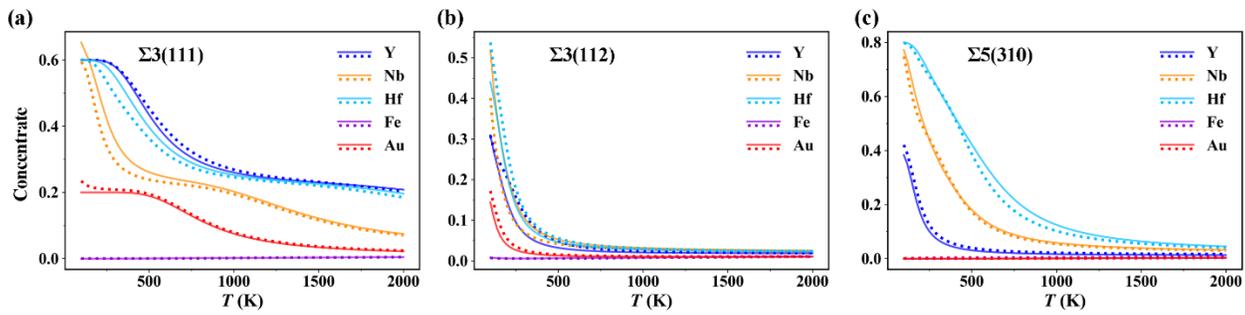

**Fig. 5 The predicted and DFT-calculated grain-boundary (GB) solute concentrate curves in the W matrix based on the White-Coghlan model[45].** (**a**) Σ3(111) GB, (**b**) Σ3(112) GB and (**c**) Σ5(310) GB. The solid curves are obtained with the segregation energies predicted from our model and the dotted lines are obtained with the DFT-calculated segregation energies.